\documentclass[ reprint,superscriptaddress, 
amsmath,amssymb,aps,longbibliography]{revtex4-1}
\pdfoutput=1
\usepackage{graphicx}
\usepackage{epstopdf}
\usepackage{color}
\usepackage{dcolumn}
\usepackage{bm}
\usepackage{hyperref}
\usepackage{upgreek}
\usepackage{units}
\usepackage{booktabs}

\newcommand{\nPhot}{\ensuremath{n_{\upgamma}}}
\newcommand{\nElec}{\ensuremath{n_{\mathrm{e}}}}
\newcommand{\SOnec}{\ensuremath{\textsl{S1}}}
\newcommand{\STwoc}{\ensuremath{\textsl{S2}}}
\newcommand{\GeVmass}{\ensuremath{\mathrm{GeV}\,c^{-2}}}

\newcommand{\ra}[1]{\renewcommand{\arraystretch}{#1}} 

\begin{document}

\title{Improved Limits on Scattering of Weakly Interacting Massive 
Particles from Reanalysis of 2013 LUX data}

\author{D.S.~Akerib} 
\affiliation{Case Western Reserve University, Department of Physics, 
10900 Euclid Ave, Cleveland, Ohio 44106, USA}
\affiliation{SLAC National Accelerator Laboratory, 2575 Sand Hill 
Road, Menlo Park, California 94205, USA}
\affiliation{Kavli Institute for Particle Astrophysics and Cosmology, 
Stanford University, 452 Lomita Mall, Stanford, California 94309, USA}

\author{H.M.~Ara\'{u}jo} 
\affiliation{Imperial College London, High Energy Physics, Blackett 
Laboratory, London SW7 2BZ, United Kingdom}

\author{X.~Bai} 
\affiliation{South Dakota School of Mines and Technology, 501 East St 
Joseph Street, Rapid City, South Dakota 57701, USA}

\author{A.J.~Bailey} 
\affiliation{Imperial College London, High Energy Physics, Blackett 
Laboratory, London SW7 2BZ, United Kingdom}

\author{J.~Balajthy} %
\affiliation{University of Maryland, Department of Physics, College 
Park, Maryland 20742, USA}

\author{P.~Beltrame} %
\affiliation{SUPA, School of Physics and Astronomy, University of 
Edinburgh, Edinburgh EH9 3FD, United Kingdom}

\author{E.P.~Bernard} %
\affiliation{Yale University, Department of Physics, 217 Prospect 
Street, 
New Haven, Connecticut 06511, USA}

\author{A.~Bernstein} %
\affiliation{Lawrence Livermore National Laboratory, 7000 East Avenue, 
Livermore, California 94551, USA}

\author{T.P.~Biesiadzinski} %
\affiliation{Case Western Reserve University, Department of Physics, 
10900 Euclid Ave, Cleveland, Ohio 44106, USA}
\affiliation{SLAC National Accelerator Laboratory, 2575 Sand Hill 
Road, Menlo Park, California 94205, USA}
\affiliation{Kavli Institute for Particle Astrophysics and Cosmology, 
Stanford University, 452 Lomita Mall, Stanford, California 94309, USA}

\author{E.M.~Boulton} %
\affiliation{Yale University, Department of Physics, 217 Prospect 
Street, 
New Haven, Connecticut 06511, USA}

% ...

\collaboration{LUX Collaboration}%

\date{\today}%

\begin{abstract}
We present constraints on  weakly interacting massive particles 
(WIMP)-nucleus scattering from the 2013 data of the Large Underground 
Xenon dark matter experiment, including $1.4\times10^{4}\;\mathrm{kg\; 
day}$ of search exposure. This new analysis incorporates several 
advances: single-photon calibration at the scintillation wavelength, 
improved event-reconstruction algorithms, a revised background model 
including events originating on the detector walls in an enlarged 
fiducial volume, and new calibrations from decays of an injected 
tritium $\upbeta$ source and from kinematically constrained nuclear 
recoils down to 1.1~keV. Sensitivity, especially to low-mass WIMPs, is 
enhanced compared to our previous results which modeled the signal only 
above a 3~keV minimum energy. Under standard dark matter halo 
assumptions and in the mass range above 4~$\mathrm{GeV}\,c^{-2}$, these 
new results give the most stringent direct limits on the 
spin-independent WIMP-nucleon cross section. The 90\% C.L.\ upper limit 
has a minimum of 0.6~zb at 33~\GeVmass\ WIMP mass.
\end{abstract}

\maketitle

Consistent evidence from a range of astrophysical observations 
suggests that cold dark matter is the dominant form of matter in our 
Galaxy and in the Universe overall 
\cite{Read:2014,Harvey:2015,Ade:2015}. Weakly interacting massive 
particles (WIMPs) are a generic class of dark matter candidate and may 
be detectable via weak-force-mediated nuclear recoils in detectors on 
Earth \cite{Goodman:1985,Feng:2010}. In October 2013, the LUX 
collaboration reported results from a 85.3 live-day exposure of a 118 
kg fiducial mass~\cite{Akerib:2013:run3}. These remain the strongest 
constraints on the spin-independent WIMP-nucleon cross section over a 
wide range of WIMP mass. They were, however, determined under the 
pessimistic assumption of zero efficiency for nuclear recoil (NR) 
events below 3~keV, which was the minimum energy at which liquid xenon 
had been calibrated at that time. Here, we present a new analysis of 
the data reported in \cite{Akerib:2013:run3} which accounts for the 
recent \textit{in situ} calibration of NR energies well below 3~keV. 
Event reconstruction and models of background are improved, and a 
further 10 days of exposure are also added. Together, these updates 
greatly enhance sensitivity to low-mass WIMPs, exploring a new region 
of dark matter parameter space.

LUX (Large Underground Xenon) is a dual-phase xenon time-projection 
chamber (TPC) with 250~kg of active liquid mass, designed to observe 
WIMPs in the local halo scattering on xenon nuclei. Energy thus 
deposited creates a primary scintillation signal, called S1, and 
ionization charge which drifts vertically in an electric field to 
produce an electroluminescence signal in the gas phase, called S2. Both 
signals are detected by photomultiplier tubes (PMTs), 61 viewing the 
TPC from above and 61 from below. A description of the detector and 
its deployment at the Sanford Underground Research Facility can be 
found in \cite{Akerib:2012:det}.

This update includes several refinements to the initial data 
processing, whereby PMT waveforms are calibrated in units of detected 
photons (phd). The pulse area estimation was further improved to reduce 
the impact of two small systematic effects. A coherent noise artifact 
consistently appeared in some channels and is now subtracted. This 
correction to each S1 or S2 pulse ranged from 0 to 0.2~phd per channel. 
The baseline estimates of the data-acquisition firmware were also found 
to introduce a small arithmetic-truncation error which was corrected. 
The mean waveform area of one detected photon within each PMT is 
calibrated using a sample of S1s below 10 phd total and near the 
detector center, after a  $<$5\% correction for photon pileup. A 
separate single-photon measurement is made using the 
electroluminescence light of single electrons (SEs). The mean over all 
PMTs agrees within 2.5\% between the two measurements. Compared to a 
previous calibration using pulsed 440~nm LEDs, these xenon light 
methods avoid pulser cross talk, avoid systematic error from assumed 
distributions by using sample means rather than parametric fits, and 
automatically account for wavelength-dependent double-photoelectron 
emission by single photons at the photocathode \cite{Faham:2015}.

Candidate single-scatter active-region events are termed ``golden'', 
and consist of one S2 preceded by one S1. S1 light in the WIMP region 
of interest is quantified using both calibrated pulse areas and pulse 
counting, whereby candidate single photons (``spikes'') are identified 
in sparse waveforms. In addition to photon statistics, pulse areas 
include fluctuations due to gain variance and single- versus 
double-photoelectron emission at the photocathode.  Therefore, counting 
discrete waveform spikes can give a more precise scintillation 
measurement over using integrated pulse areas. A parametrization of the 
maximum-likelihood number of photons, as a function of area and spike 
count, is computed from simulated pileup in time and measured photon 
area distributions. For S1s above 20 keV electron-recoil (ER) 
equivalent energy and for all S2s, where pileup is prevalent, detected 
photons are estimated using pulse area alone. The drift time between S1 
and S2 gives the vertical location of each event to millimeter 
precision ($\sigma=0.9\;\textrm{mm}$ measured with coincident Bi-Po 
decays \cite{Faham:2014}). S2 positions in the $x$-$y$ plane are 
estimated using data-derived parametrizations of individual top-array 
PMT responses \cite{Solovov:2012}. The gate and cathode electrode grids 
establish a field, with a mean and range in the fiducial volume of 
$180\pm20$~$\mathrm{V\,cm}^{-1}$, to drift charge from the active 
volume towards the liquid surface. The field is nonuniform due to 
geometric effects similar to \cite{Mei:2011}. A weak radial component 
moves drifting electrons inwards from the site of ionization by up to 
4.6~cm for the outer bottom edge of the fiducial volume, in agreement 
with an electrostatic model of the drift field 
\cite{Akerib:2015:comprehensive}. We account for this effect by 
exploiting the spatial uniformity of a $^{\mathrm{83m}}$Kr calibration 
source \cite{Kastens:2009,Manalaysay:2010} to derive a mapping between 
S2 and vertex position. Position variables used in later analysis refer 
to the reconstructed vertex: the standard deviations of the 
reconstructed $x$ and $y$ have a statistical contribution of 10~mm at 
the S2 threshold, and a 5~mm systematic contribution estimated from the 
reconstruction of the chamber walls and of a collimated neutron beam 
\cite{Akerib:2016:dd}.

Weekly calibrations with the monoenergetic $^{\mathrm{83m}}$Kr source 
are used to derive, from the estimates of detected photons and event 
position, two corrected variables, called \SOnec\ and \STwoc, which 
equalize detector response throughout the active volume. They are 
proportional, respectively, to the scintillation light and ionization 
charge leaving the interaction site. By convention, \SOnec\ equals the 
raw number of detected photons for events at the center of the 
detector. Similarly, events at the center would, in the absence of 
signal charge loss to impurities during drifting, have a mean of 
\STwoc\ detected S2 photons. Calibration relative to these reference 
points accounts for position dependence in the efficiency to extract an 
electron into the gas, electroluminescence yield, and photon-detection 
efficiency, and for time-dependent xenon purity. In 
\cite{Akerib:2013:run3}, ionization was estimated using only the bottom 
PMT array, over which S2 light is quite uniform. However, a subsequent 
large-sample calibration with a dissolved tritiated methane source 
\cite{Akerib:2015:tritium} has demonstrated that using all PMTs reduces 
by 20\% the rate of leakage ER events below the Gaussian mean 
$\log(\STwoc/\SOnec)$ of NR calibration at a given \SOnec. We find 
that, after flat fielding, the reduced variance from measuring more 
photons outweighs residual nonuniformity in the top array response. 
The sum of top and bottom arrays is thus adopted for \STwoc.

\begin{figure}[t]
\begin{center}
\includegraphics[width=0.5\textwidth,clip]{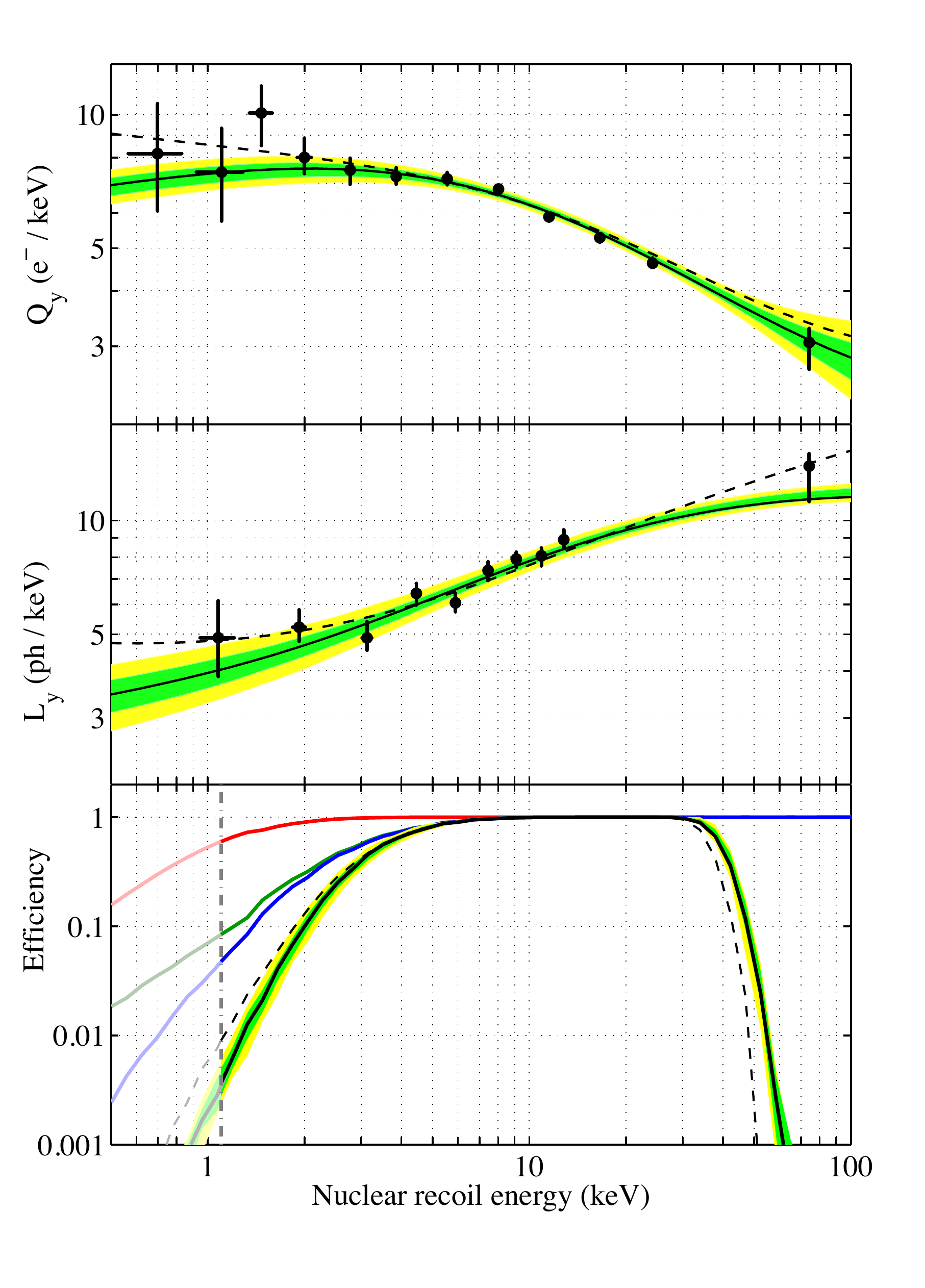}
\caption{\label{fig:nr_efficiency} \textit{Top, middle:} yields of 
electrons and photons, respectively, for nuclear recoils in LUX, 
measured \textit{in situ} with D-D neutrons. Error bars are 
statistical. \textit{Bottom:} efficiencies for NR event detection, 
averaged over the fiducial volume and estimated using LUXS\textsc{im} 
with parameters tuned to D-D calibration. In descending order of 
efficiency---red: detection of an S2 ($\geq$2 electrons emitted); 
green: detection of an S1 ($\geq$2 PMTs detecting photons); blue: 
detection of both an S1 and an S2; black: detection passing thresholds 
in \SOnec\ and raw S2 size. The $97.5\%\pm1.7\%$ event-classification 
efficiency is applied as an additional, energy-independent scaling. The 
vertical line at 1.1~keV marks the low-energy cutoff applied in the 
signal model. \textit{All panels:} solid lines show the best fit of the 
Lindhard parametrization; shaded regions span its 1- and 2-$\sigma$ 
uncertainty used for the final result. Dashed lines show the best fit 
of the alternate, Bezrukov NR parametrization.}
\end{center}
\end{figure}

The detector-specific gain factors $g_\textsc{1}$ and $g_\textsc{2}$ 
are defined via the expectation values \mbox{$\langle 
\SOnec\rangle=g_\textsc{1}\nPhot$} and 
\mbox{$\langle\STwoc\rangle=g_\textsc{2}\nElec$}, given \nPhot\ initial 
photons and \nElec\ initial electrons leaving the interaction site. 
Their values in LUX were obtained by the technique of \cite{Doke:2002} 
using a set of monoenergetic electron-recoil sources as in 
\cite{Phelps:2014}. The sum of the photon yield and the electron yield 
is observed to be constant with energy, equal to the reciprocal of the 
$W$~value as defined in \cite{Shutt:2007}; however, the individual 
yields do vary, because charge recombination probability depends upon 
energy, $E$. In a plot of $\STwoc/E$ versus $\SOnec/E$, the sources 
trace a line and a fit to this line measures the gain factors: 
$g_\textsc{1}=\left(0.117\pm0.003\right)$~phd per photon and 
$g_\textsc{2}=\left(12.1\pm0.8\right)$~phd per electron, with 
anticorrelation \mbox{$\rho=-0.6$}. Calibrating \SOnec\ and 
$g_\textsc{1}$ in units of detected VUV photons results in a numerical 
shift relative to the previous, smaller units of photoelectrons (phe) 
but is preferred because $g_\textsc{1}$ thus defined is the probability 
for an initial photon to cause a detectable PMT response. Using yields 
at many discrete energies is also more robust than the single spectral 
fit used to estimate values of 
$g_\textsc{1}=\left(0.14\pm0.01\right)$~phe per photon and 
$g_\textsc{2}=\left(16.0\pm0.3\right)$~phe per electron in 
\cite{Akerib:2013:run3}.

The fiducial range in drift time, mitigating radiogenic backgrounds 
from detector materials, is unchanged from \cite{Akerib:2013:run3} at 
38--305~$\upmu$s (48.6--8.5~cm above the faces of the bottom PMTs in 
$z$). A data-driven model of events originating on detector sidewalls 
allows a larger fiducial radius of 20~cm. The fiducial mass was 
measured as a fraction of the known active xenon mass by counting 
tritium events: the result of (145.4$\pm$1.3)~kg is consistent with 
the 147~kg expected from geometry. S1 pulses are required to have 
two-PMT coincidence and \SOnec\ in the range 1--50~phd. Normalizing to 
the detector center means that \SOnec\ can be below 2.0~phd even with 
two photons detected. A lower analysis threshold of 165~phd raw S2 
size (6.7 times the mean SE response) is applied to mitigate the 
random coincidence background from smaller, isolated S2s.

The LUX NR response in \STwoc\ and \SOnec\ has been measured 
\textit{in situ} using monoenergetic neutrons from an Adelphi DD108 
deuterium-deuterium (D-D) fusion source. The yields are presented in 
Fig.~\ref{fig:nr_efficiency}. The dominant systematics in these charge 
and light calibrations correspond to a uniform 9\% and 3\%, 
respectively \cite{Malling:2014,Verbus:2015,Akerib:2016:dd}. The NR 
response in \STwoc\ was measured with an absolute determination of the 
deposited energy from scattering angles in multiple-vertex events. 
This calibration of the NR signal yields directly improves sensitivity 
to low-mass WIMPs over \cite{Akerib:2013:run3}.

To compute WIMP signal probability density functions (PDFs) from the 
D-D calibration and account for uncertainty, an empirical response 
model was fitted simultaneously to the yields and to the median 
\STwoc\ versus \SOnec\ of single-scatter NR events. The mean fraction 
of recoil energy lost to electrons, $ {\cal{L}}\left(E\right)$, is 
described by the Lindhard model \cite{Lindhard:1963}. Scintillation and 
ionization quanta leaving the track are described by an 
energy-independent ratio of initial excitons and ions, followed by 
charge recombination according to the Thomas-Imel box model 
\cite{Thomas:1987} and biexcitonic quenching including Penning 
ionization \cite{Hitachi:1992,Mei:2008}. \SOnec\ and \STwoc\ are then 
generated via standard statistical distributions which model stages of 
detector response (collection of scintillation photons, attenuation of 
the ionization signal before S2 production, photoelectron and SE 
distributions). The full model is described in \cite{Akerib:2016:dd} 
and the fit procedure follows \cite{Lenardo:2014}. An alternate 
parametrization of ${\cal{L}}\left(E\right)$ by Bezrukov \emph{et 
al.}\ 
\cite{Bezrukov:2011} is similarly consistent with calibration data and 
implies higher signal efficiency at low energies; it is shown for 
reference but does not enter into the reported limit.  
Figure~\ref{fig:nr_efficiency} shows the best fits to experimental 
yields of signal quanta for both parametrizations.

Nuclear-recoil energy spectra for the WIMP signal are derived from a 
standard Maxwellian velocity distribution with $v_{0}$ = 220~km/s, 
$v_{\mathrm{esc}}$ = 544~km/s, $\rho_{0}$ = 0.3~GeV/cm$^{3}$, average 
Earth velocity during data taking of 245~km/s, and a Helm form factor, 
as in \cite{Akerib:2013:run3}. Following the same criterion as that 
analysis, but with new calibration data, the signal spectrum is 
assumed zero below the lowest \mbox{D-D} \SOnec\ calibration point of 
1.1~keV. Signal PDFs and rates as a function of the spin-independent 
WIMP-nucleon cross section, $\sigma_{n}$, are computed from the 
empirical NR response model. Uncertainties in the absolute values of 
$g_\textsc{1}$ and $g_\textsc{2}$ do not propagate to the signal model, 
because it is calibrated \textit{in situ} in the \SOnec\ and \STwoc\ 
variables. The non-negligible signal-model uncertainties are 
incorporated in the likelihood via two nuisance parameters with 
Gaussian constraints from the \mbox{D-D} calibration (see Table 
\ref{tab:fit_parameters}): the Lindhard $k$ parameter and the S2 gain 
during \mbox{D-D} calibration in November 2013 relative to the WIMP 
search, $g_{2,\textsc{dd}}/g_{2,\textsc{ws}}$.

The efficiency for WIMP-nuclear recoils to appear as events in the 
search data is the product of several detection stages. Modeling the 
WIMP signal only above 1.1~keV includes 0.3\% of the recoil spectrum 
for a 4~\GeVmass\ WIMP, rising to 94\% in the high-mass limit. The 
efficiency to generate an S1 and an S2 passing all analysis thresholds 
in the best-fit NR model, shown along with systematic variations in 
Fig.~\ref{fig:nr_efficiency}, rises from 0.3\% at the 1.1~keV cutoff to 
50\% at 3.3~keV. Finally, identification of S1 and S2 within real 
waveforms can fail in ways not reproduced by simulation, for instance 
where the hit-pattern or pulse-shape variables used in classification 
are biased by PMT afterpulsing. The probability to thus discard events 
was found by visually inspecting 4000 AmBe calibration events: the 
pulse-identification efficiency for events in the WIMP region of 
interest and passing the analysis thresholds was found to be 
$97.5\%\pm1.7\%$, and is implemented as an energy-independent scaling.

\begin{table}[t]
\ra{1.2}
\caption{\label{tab:fit_parameters}Nuisance parameters in the global 
best fit to 95-day search data. Constraints are Gaussian with means 
and standard deviations indicated. Event counts are after cuts and 
analysis thresholds. The best-fit model has zero contribution from the 
signal PDF. In this case the signal-model parameters simply float to 
the central values of their constraints, and so are not listed.}
\begin{tabular}{@{}lcc@{}}
Parameter & Constraint & Fit value\\ 
\hline
Lindhard $k$ & $0.174\pm0.006$ & $\cdots$ \\
S2 gain ratio: $g_{2,\textsc{dd}}/g_{2,\textsc{ws}}$ & $0.94\pm0.04$ & 
$\cdots$ \\ 
Low-$z$-origin $\upgamma$ counts: $\mu_{\upgamma,\mathrm{bottom}}$ & 
$172\pm74$ & $165\pm16$ \\
Other $\upgamma$ counts: $\mu_{\upgamma,\mathrm{rest}}$ & $247\pm106$ 
& $228\pm19$ \\
$\upbeta$ counts: $\mu_{\mathrm{\upbeta}}$ & $55\pm22$ & $84\pm15$ \\
$^{127}$Xe counts: $\mu_{\textrm{Xe-127}}$ & $91\pm27$ & $78\pm12$ \\
$^{37}$Ar counts: $\mu_{\textrm{Ar-37}}$ & $\cdots$ & $12\pm 8$ \\
Wall counts: $\mu_{\mathrm{wall}}$ & $24\pm7$ & $22\pm4$ \\
\hline
\end{tabular}
\end{table}

Radiogenic backgrounds are again estimated as in 
\cite{Akerib:2015:bg}, but with the revised data-reduction techniques 
and cuts. The added acceptance increases the expected neutron 
background to $0.08\pm0.01$ NR events in the WIMP-search sample. Random 
coincidence of isolated S1s (having rate $1\;\mathrm{s}^{-1}$) and S2s 
($5\times10^{-4}\;\mathrm{s}^{-1}$) within a physical drift time 
causes an expected 1.1 events in the full search range of \SOnec\ and 
\STwoc. Coherent neutrino-nucleus scattering by $^{8}$B solar 
neutrinos contributes 0.10 (0.16) golden events under the Lindhard 
(Bezrukov) yield model. None of these small background populations are 
included in the model.

Isolated low-energy ER events in the fiducial volume arise from four 
sources: Compton scattering of $\upgamma$~rays from detector component 
radioactivity, $^{85}$Kr or Rn-daughter contaminants in the liquid 
undergoing $\upbeta$ decay with no accompanying $\upgamma$~rays 
detected, x~rays following those $^{127}$Xe electron-capture decays 
where the coincident $\upgamma$~ray escapes the xenon, and a line at 
2.8~keV, evident due to the improved energy resolution and consistent 
with electron-capture decays in the fiducial volume by $^{37}$Ar 
nuclei. Measurements of the $^{37}$Ar concentration in lab air are 
planned and will, together with limits on air leaks from xenon 
sampling results, give an upper limit on rate; it is currently an 
unconstrained fit parameter.

The G\textsc{eant}4-based LUXS\textsc{im} package, incorporating the 
NEST model for 
signal generation in the xenon 
\cite{Agostinelli:2003,Szydagis:2011,Szydagis:2013,Akerib:2012:luxsim},
 was tuned to the \SOnec-\STwoc\ distribution of $1.8\times10^{5}$ 
fiducial-volume electron recoils from the internal tritium source. 
Good agreement was obtained from threshold to the 18.6~keV end point, 
well above the WIMP signal in both light and charge, and the 
reconstructed $\upbeta$ spectrum validates the $g_\textsc{1}$ and 
$g_\textsc{2}$ values measured with line sources 
\cite{Akerib:2015:tritium}. 
Simulated waveforms, processed with the same data-reduction software 
and event selection as applied to the search data, are used to model 
the ER backgrounds in \SOnec\ and \STwoc.

Events due to detector component radioactivity, both within and above 
the energy region of interest, were simulated with LUXS\textsc{im}. 
The 
high-energy spectral agreement between data and simulation based on 
$\upgamma$ screening is generally good \cite{Malling:2014, 
Akerib:2015:bg}; however, we observe an excess of ER events with 
500--1500 keV energy concentrated in the lowest 10~cm of the active 
region. Its precise origin is unknown but the spectrum can be 
reproduced by simulating additional, heavily downscattered 
$^{238}$U~chain, $^{232}$Th~chain, and $^{60}$Co $\upgamma$~rays in 
the center of a large copper block below the PMTs. This implies an 
extra 105 low-energy Compton-scatter events, included in the 
background model. The $\upgamma$-ray population is subdivided into two 
spatial distributions with floating normalization: one generated by the 
bottom PMT array, its support structure, and the bottom $\upgamma$-ray 
shield; and one from the rest of the detector.

A final source of background, newly modeled here, is the tail in 
reconstructed $r$ of events on the PTFE sidewalls. The \SOnec-\STwoc\ 
distribution of background events on the walls differs from that in 
the liquid bulk. Charge collection is incomplete, so the ER population 
extends to lower values of \STwoc. There are, in addition, true 
nuclear recoils from the daughter $^{206}$Pb nuclei of $\upalpha$ decay 
by $^{210}$Po plated on the wall. The leakage of wall events towards 
smaller $r$ depends strongly, via position resolution, on S2 size. The 
wall population in the fiducial volume thus appears close to the S2 
threshold, largely below the signal population in \STwoc\ at given 
\SOnec. It is modeled empirically using high-$r$ and low-\STwoc\ 
sidebands in the search data \cite{Lee:2015}.

Systematic uncertainties in background rates are treated via nuisance 
parameters in the likelihood: their constraints are listed with other 
fit parameters in Table~\ref{tab:fit_parameters}. \SOnec, \STwoc, $z$, 
and $r$ are each useful discriminants against backgrounds and cross 
sections are tested via the likelihood of the search events in these 
four observables.
 
\begin{figure}[t]
\begin{center}
\includegraphics[width=0.48\textwidth,clip]{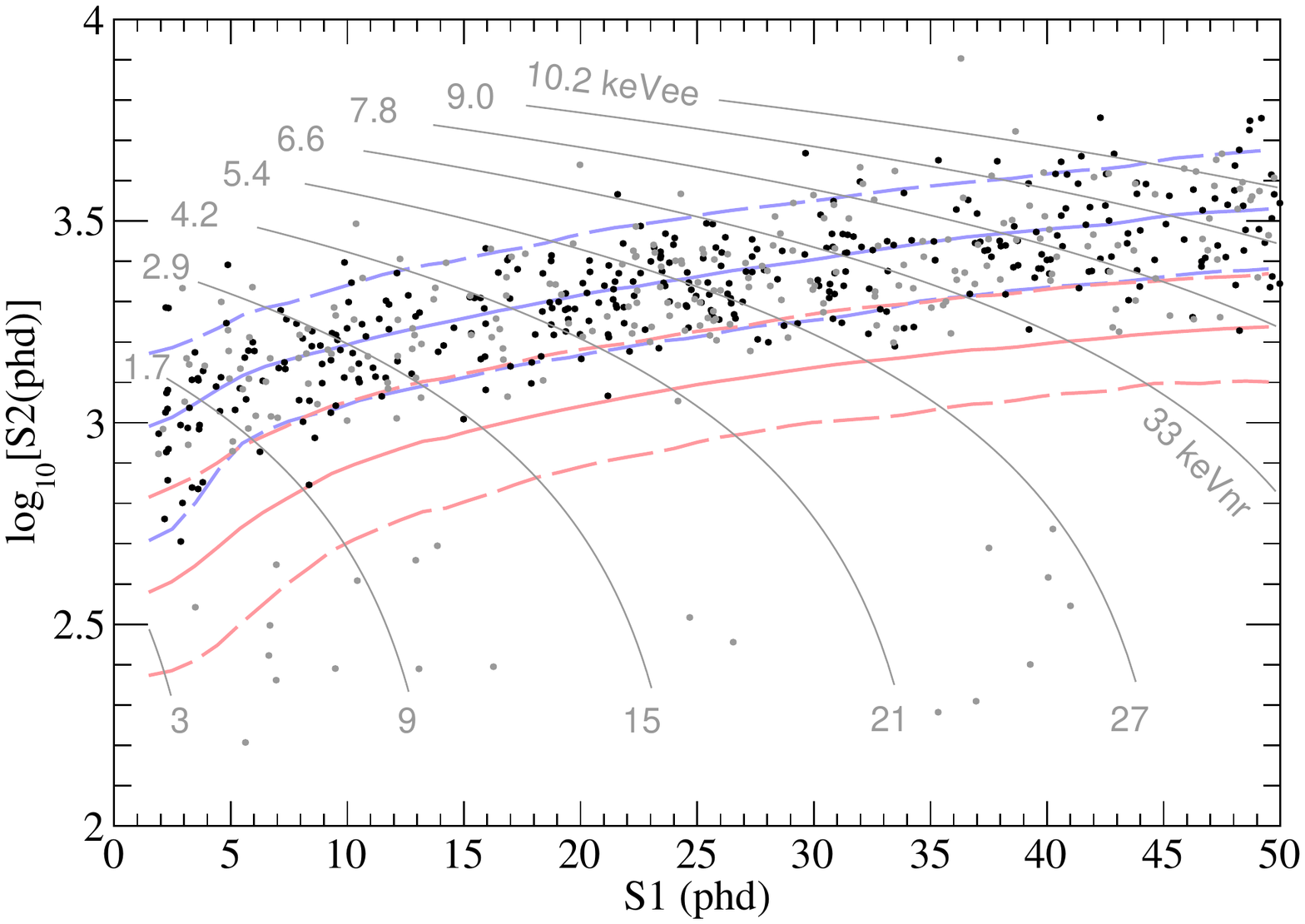}
\caption{\label{fig:WS_scatterplot} Observed events in the 2013 LUX  
exposure of 95 live days and 145~kg fiducial mass. Points at $<$18~cm 
radius are black; those at 18--20~cm are gray. Distributions of 
uniform-in-energy electron recoils (blue) and an example 
50~$\mathrm{GeV}\,c^{-2}$ WIMP signal (red) are indicated by 50th 
(solid), 10th, and 90th  (dashed) percentiles of \STwoc\ at given 
\SOnec. Gray lines, with ER scale of keVee at top and Lindhard-model 
NR scale of keVnr at bottom, are contours of the linear combined 
\SOnec-and-\STwoc\ energy estimator \cite{Shutt:2007}.}
\end{center}
\end{figure}

Search data were acquired between April 24th and September 1st, 2013. 
Two classes of cuts based on prevailing detector conditions assure 
well-measured events in both low-energy calibration and WIMP-search 
samples. Firstly, data taken during excursions in macroscopic detector 
properties, such as xenon circulation outages or instability of applied 
high voltage, are removed, constituting 0.8\% of gross live time. 
Secondly, an upper threshold is imposed on summed pulse area during the 
event window but outside S1 and S2. It removes triggers during the 
aftermath of photoionization and delayed electron emission following 
large S2s. The threshold is set for $>$99\% tritium acceptance and 
removes 1\% of gross live time \cite{Chapman:2014}. We report on 95.0 
live days. Figure~\ref{fig:WS_scatterplot} shows the measured light and 
charge of the 591 surviving events in the fiducial volume.

\begin{figure}[t]
\begin{center}
\includegraphics[width=0.48\textwidth,clip]{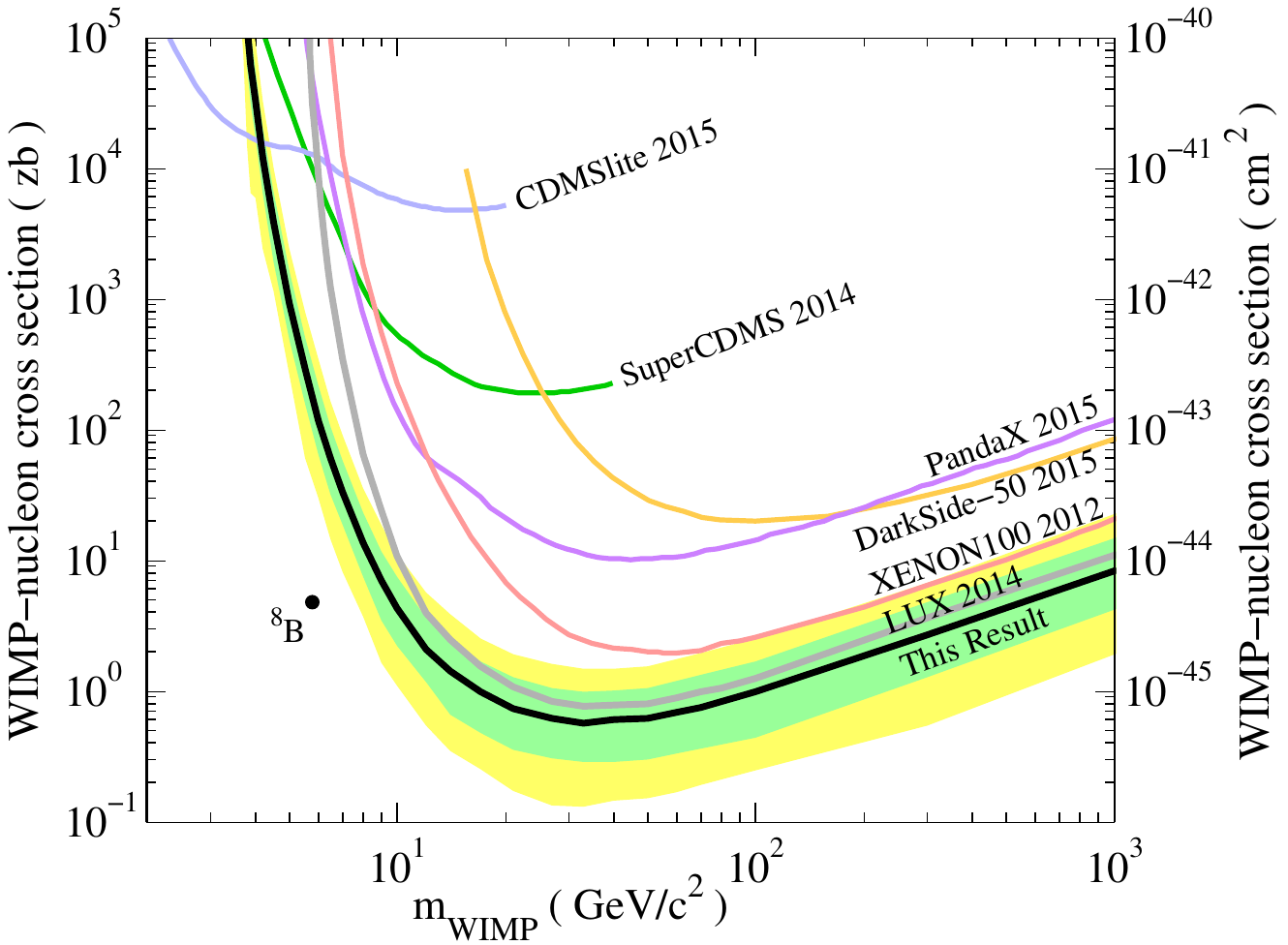}
\caption{\label{fig:limit} Upper limits on the spin-independent 
elastic 
WIMP-nucleon cross section at 90\% C.L. Observed limit in black, with 
the 1- and 2-$\sigma$ ranges of background-only trials shaded green 
and 
yellow.  Also shown are limits from the first LUX analysis 
\cite{Akerib:2013:run3} (gray), SuperCDMS \cite{Scdms:2014} (green), 
CDMSlite \cite{Cdmslite:2015} (light blue), XENON100 
\cite{Xenon100:2012} (red), DarkSide-50 \cite{Darkside:2015} (orange), 
and PandaX \cite{Pandax:2015} (purple).  The expected spectrum of 
coherent neutrino-nucleus scattering by $^{8}$B solar neutrinos can be 
fit by a WIMP model as in \cite{Billard:2013}, plotted here as a black 
dot.}
\end{center}
\end{figure}

A double-sided, profile-likelihood-ratio (PLR) statistic 
\cite{Cowan:2011} is employed to test signal hypotheses.
For each WIMP mass we scan over cross section to construct a 90\% 
confidence interval, with test statistic distributions evaluated by MC 
using the R\textsc{oo}S\textsc{tats} package \cite{Moneta:2010}. At all 
masses, the maximum-likelihood value of $\sigma_n$ is found to be zero. 
The background-only model gives a good fit to the data, with KS test 
$p$~values of 0.05, 0.07, 0.34, and 0.64 for the projected 
distributions in \SOnec, \STwoc, $r$, and $z$ respectively. Upper 
limits on cross section for WIMP masses from 4 to 1000~\GeVmass\ are 
shown in Fig.~\ref{fig:limit}; above, the limit increases in proportion 
to mass until ${\gtrsim}10^8\;\GeVmass$, $10^{6}\;\textrm{zb}$, where 
the Earth begins to attenuate the WIMP flux. The raw PLR result lies 
between one and two Gaussian $\sigma$ below the expected limit from 
background trials. We apply a power constraint \cite{Cowan:2011:pcl} at 
the median so as not to exclude cross sections for which sensitivity is 
low through chance background fluctuation. We include systematic 
uncertainties in the nuclear recoil response in the PLR, which has a 
modest effect on the limit with respect to assuming the best-fit model 
exactly: less than 20\% at all masses.  Limits calculated with the 
alternate, Bezrukov parametrization would be 0.48, 1.02, and 1.05 times 
the reported ones at 4, 33, and 1000~\GeVmass, respectively. 
Uncertainties in the assumed dark matter halo are beyond the scope of 
this Letter but are reviewed in, e.g., \cite{McCabe:2010}. Limits on 
spin-dependent cross sections are presented elsewhere 
\cite{Akerib:2016:sd}.

In conclusion, reanalysis of the 2013 LUX data has excluded new WIMP 
parameter space. The added fiducial mass and live time, and better 
resolution of light and charge yield a 23\% improvement in sensitivity 
at high WIMP masses over the first LUX result. The reduced, 1.1~keV 
cutoff in the signal model improves sensitivity by 2\% at high masses 
but is the dominant effect below 20~\GeVmass, and the range 5.2 to 
3.3~\GeVmass\ is newly demonstrated to be detectable in xenon. These 
techniques further enhance the prospects for discovery in the ongoing 
300-day LUX search and the future LUX-ZEPLIN \cite{Akerib:2015:cdr} 
experiment.

This work was partially supported by the U.S. Department of Energy 
(DOE) under Awards No.\ DE-FG02-08ER41549, No.\ DE-FG02-91ER40688, No.\ 
DE-FG02-95ER40917, No.\ DE-FG02-91ER40674, No.\ DE-NA0000979, No.\ 
DE-FG02-11ER41738, No.\ DE-SC0006605, No.\ DE-AC02-05CH11231, No.\ 
DE-AC52-07NA27344, and No.\ DE-FG01-91ER40618; the U.S. National 
Science Foundation under Grants  No.\ PHYS-0750671, No.\ PHY-0801536, 
No.\ PHY-1004661, No.\ PHY-1102470, No.\ PHY-1003660, No.\ PHY-1312561, 
No.\ PHY-1347449, and No.\ PHY-1505868; the Research Corporation Grant 
No.\ RA0350; the Center for Ultra-low Background Experiments in the 
Dakotas (CUBED); and the South Dakota School of Mines and Technology 
(SDSMT). LIP-Coimbra acknowledges funding from Funda\c{c}\~{a}o para a 
Ci\^{e}ncia e a Tecnologia (FCT) through the Project-Grant No.\ 
PTDC/FIS-NUC/1525/2014. Imperial College and Brown University thank the 
UK Royal Society for travel funds under the International Exchange 
Scheme (IE120804). The UK groups acknowledge institutional support from 
Imperial College London, University College London and Edinburgh 
University, and from the Science \& Technology Facilities Council for 
PhD studentships No.\ ST/K502042/1 (AB), No.\ ST/K502406/1 (SS), and 
No.\ ST/M503538/1 (KY). The University of Edinburgh is a charitable 
body, registered in Scotland, with Registration No.\ SC005336. This 
research was conducted using computational resources and services at 
the Center for Computation and Visualization, Brown University. We 
gratefully acknowledge the logistical and technical support and the 
access to laboratory infrastructure provided to us by the Sanford 
Underground Research Facility (SURF) and its personnel at Lead, South 
Dakota. SURF was developed by the South Dakota Science and Technology 
Authority, with an important philanthropic donation from T. Denny 
Sanford, and is operated by Lawrence Berkeley National Laboratory for 
the Department of Energy, Office of High Energy Physics. We thank Felix 
Kahlhoefer and Sebastien Wild for uncovering a mistake in a preprint of 
this work. 
\bibliography{main}

 \newcommand{\noop}[1]{}
\begin{thebibliography}{46}%
\makeatletter
\providecommand \@ifxundefined [1]{%
 \@ifx{#1\undefined}
}%
\providecommand \@ifnum [1]{%
 \ifnum #1\expandafter \@firstoftwo
 \else \expandafter \@secondoftwo
 \fi
}%
\providecommand \@ifx [1]{%
 \ifx #1\expandafter \@firstoftwo
 \else \expandafter \@secondoftwo
 \fi
}%
\providecommand \natexlab [1]{#1}%
\providecommand \emph  [1]{``#1''}%
\providecommand \bibnamefont  [1]{#1}%
\providecommand \bibfnamefont [1]{#1}%
\providecommand \citenamefont [1]{#1}%
\providecommand \href@noop [0]{\@secondoftwo}%
\providecommand \href [0]{\begingroup \@sanitize@url \@href}%
\providecommand \@href[1]{\@@startlink{#1}\@@href}%
\providecommand \@@href[1]{\endgroup#1\@@endlink}%
\providecommand \@sanitize@url [0]{\catcode `\\12\catcode `\$12\catcode
  `\&12\catcode `\#12\catcode `\^12\catcode `\_12\catcode `\%12\relax}%
\providecommand \@@startlink[1]{}%
\providecommand \@@endlink[0]{}%
\providecommand \url  [0]{\begingroup\@sanitize@url \@url }%
\providecommand \@url [1]{\endgroup\@href {#1}{\urlprefix }}%
\providecommand \urlprefix  [0]{URL }%
\providecommand \Eprint [0]{\href }%
\providecommand \doibase [0]{http://dx.doi.org/}%
\providecommand \selectlanguage [0]{\@gobble}%
\providecommand \bibinfo  [0]{\@secondoftwo}%
\providecommand \bibfield  [0]{\@secondoftwo}%
\providecommand \translation [1]{[#1]}%
\providecommand \BibitemOpen [0]{}%
\providecommand \bibitemStop [0]{}%
\providecommand \bibitemNoStop [0]{.\EOS\space}%
\providecommand \EOS [0]{\spacefactor3000\relax}%
\providecommand \BibitemShut  [1]{\csname bibitem#1\endcsname}%
\let\auto@bib@innerbib\@empty
\bibitem [{\citenamefont {Read}(2014)}]{Read:2014}%
  \BibitemOpen
  \bibfield  {author} {\bibinfo {author} {\bibfnamefont {J~I}\ 
\bibnamefont
  {Read}},\ }\bibfield  {title} {\emph {\bibinfo {title} {The local 
dark
  matter density}},\ }\href 
{http://stacks.iop.org/0954-3899/41/i=6/a=063101}
  {\bibfield  {journal} {\bibinfo  {journal} {J.\ Phys.\ G}\ }\textbf 
{\bibinfo
  {volume} {41}},\ \bibinfo {pages} {063101} (\bibinfo {year}
  {2014})}\BibitemShut {NoStop}%
\bibitem [{\citenamefont {Harvey}\ \emph {et~al.}(2015)\citenamefont 
{Harvey},
  \citenamefont {Massey}, \citenamefont {Kitching}, \citenamefont 
{Taylor},\
  and\ \citenamefont {Tittley}}]{Harvey:2015}%
  \BibitemOpen
  \bibfield  {author} {\bibinfo {author} {\bibfnamefont {David}\ 
\bibnamefont
  {Harvey}}, \bibinfo {author} {\bibfnamefont {Richard}\ \bibnamefont
  {Massey}}, \bibinfo {author} {\bibfnamefont {Thomas}\ \bibnamefont
  {Kitching}}, \bibinfo {author} {\bibfnamefont {Andy}\ \bibnamefont 
{Taylor}},
  \ and\ \bibinfo {author} {\bibfnamefont {Eric}\ \bibnamefont 
{Tittley}},\
  }\bibfield  {title} {\emph {\bibinfo {title} {The nongravitational
  interactions of dark matter in colliding galaxy clusters}},\ }\href 
{\doibase
  10.1126/science.1261381} {\bibfield  {journal} {\bibinfo  {journal}
  {Science}\ }\textbf {\bibinfo {volume} {347}},\ \bibinfo {pages} 
{1462--1465}
  (\bibinfo {year} {2015})}\BibitemShut {NoStop}%
\bibitem [{\citenamefont {Ade}\ \emph {et~al.}()\citenamefont {Ade} 
\emph
  {et~al.}}]{Ade:2015}%
  \BibitemOpen
  \bibfield  {author} {\bibinfo {author} {\bibnamefont {Ade}} \emph 
{et~al.}
  (\bibinfo {collaboration} {{Planck Collaboration}}),\ }\bibfield  
{title}
  {\emph {\bibinfo {title} {{Planck 2015 results. XIII. Cosmological
  parameters}}},\ }\href@noop {} {\ }\Eprint
  {http://arxiv.org/abs/1502.01589v2} {arXiv:1502.01589v2} \BibitemShut
  {NoStop}%
\bibitem [{\citenamefont {Goodman}\ and\ \citenamefont
  {Witten}(1985)}]{Goodman:1985}%
  \BibitemOpen
  \bibfield  {author} {\bibinfo {author} {\bibfnamefont {Mark~W.}\ 
\bibnamefont
  {Goodman}}\ and\ \bibinfo {author} {\bibfnamefont {Edward}\ 
\bibnamefont
  {Witten}},\ }\bibfield  {title} {\emph {\bibinfo {title} 
{Detectability of
  certain dark-matter candidates}},\ }\href {\doibase 
10.1103/PhysRevD.31.3059}
  {\bibfield  {journal} {\bibinfo  {journal} {Phys.\ Rev.\ D}\ }\textbf
  {\bibinfo {volume} {31}},\ \bibinfo {pages} {3059--3063} (\bibinfo 
{year}
  {1985})}\BibitemShut {NoStop}%
\bibitem [{\citenamefont {Feng}(2010)}]{Feng:2010}%
  \BibitemOpen
  \bibfield  {author} {\bibinfo {author} {\bibfnamefont {Jonathan~L.}\
  \bibnamefont {Feng}},\ }\bibfield  {title} {\emph {\bibinfo {title} 
{{Dark
  matter candidates from particle physics and methods of 
detection}}},\ }\href
  {\doibase 10.1146/annurev-astro-082708-101659} {\bibfield  {journal}
  {\bibinfo  {journal} {Annu.\ Rev.\ Astron.\ Astrophys.}\ }\textbf 
{\bibinfo
  {volume} {48}},\ \bibinfo {pages} {495--545} (\bibinfo {year}
  {2010})}\BibitemShut {NoStop}%
\bibitem [{\citenamefont {Akerib}\ \emph {et~al.}(2014)\citenamefont 
{Akerib}
  \emph {et~al.}}]{Akerib:2013:run3}%
  \BibitemOpen
  \bibfield  {author} {\bibinfo {author} {\bibfnamefont {D.S.}\ 
\bibnamefont
  {Akerib}} \emph {et~al.} (\bibinfo {collaboration} {LUX 
collaboration}),\
  }\bibfield  {title} {\emph {\bibinfo {title} {{First results from 
the LUX
  dark matter experiment at the Sanford Underground Research 
Facility}}},\
  }\href {\doibase 10.1103/PhysRevLett.112.091303} {\bibfield  
{journal}
  {\bibinfo  {journal} {Phys.\ Rev.\ Lett.}\ }\textbf {\bibinfo 
{volume}
  {112}},\ \bibinfo {pages} {091303} (\bibinfo {year} 
{2014})}\BibitemShut
  {NoStop}%
\bibitem [{\citenamefont {Akerib}\ \emph {et~al.}(2013)\citenamefont 
{Akerib}
  \emph {et~al.}}]{Akerib:2012:det}%
  \BibitemOpen
  \bibfield  {author} {\bibinfo {author} {\bibfnamefont {D.S.}\ 
\bibnamefont
  {Akerib}} \emph {et~al.} (\bibinfo {collaboration} {LUX 
Collaboration}),\
  }\bibfield  {title} {\emph {\bibinfo {title} {{The Large Underground 
Xenon
  (LUX) Experiment}}},\ }\href {\doibase 10.1016/j.nima.2012.11.135} 
{\bibfield
   {journal} {\bibinfo  {journal} {Nucl.\ Instrum.\ Methods Phys.\, 
Sect.\ A}\
  }\textbf {\bibinfo {volume} {704}},\ \bibinfo {pages} {111--126} 
(\bibinfo
  {year} {2013})}\BibitemShut {NoStop}%
\bibitem [{\citenamefont {Faham}\ \emph {et~al.}(2015)\citenamefont 
{Faham},
  \citenamefont {Gehman}, \citenamefont {Currie}, \citenamefont {Dobi},
  \citenamefont {Sorensen},\ and\ \citenamefont 
{Gaitskell}}]{Faham:2015}%
  \BibitemOpen
  \bibfield  {author} {\bibinfo {author} {\bibfnamefont {C.H.}\ 
\bibnamefont
  {Faham}}, \bibinfo {author} {\bibfnamefont {V.M.}\ \bibnamefont 
{Gehman}},
  \bibinfo {author} {\bibfnamefont {A.}~\bibnamefont {Currie}}, 
\bibinfo
  {author} {\bibfnamefont {A.}~\bibnamefont {Dobi}}, \bibinfo {author}
  {\bibfnamefont {P.}~\bibnamefont {Sorensen}}, \ and\ \bibinfo 
{author}
  {\bibfnamefont {R.J.}\ \bibnamefont {Gaitskell}},\ }\bibfield  
{title}
  {\emph {\bibinfo {title} {{Measurements of wavelength-dependent 
double
  photoelectron emission from single photons in VUV-sensitive 
photomultiplier
  tubes}}},\ }\href {\doibase 10.1088/1748-0221/2015/9/P09010} 
{\bibfield
  {journal} {\bibinfo  {journal} {J.\ Inst.}\ }\textbf {\bibinfo 
{volume}
  {10}},\ \bibinfo {eid} {P09010} (\bibinfo {year} {2015})}\BibitemShut
  {NoStop}%
\bibitem [{\citenamefont {Faham}(2014)}]{Faham:2014}%
  \BibitemOpen
  \bibfield  {author} {\bibinfo {author} {\bibfnamefont {C.H.}\ 
\bibnamefont
  {Faham}},\ }\emph {\bibinfo {title} {Prototype, surface 
commissioning and
  photomultiplier tube characterization for the Large Underground 
Xenon (LUX)
  direct dark matter search experiment}},\ \href
  {https://repository.library.brown.edu/studio/item/bdr:386139/} {Ph.D.
  thesis},\ \bibinfo  {school} {Brown University} (\bibinfo {year}
  {2014})\BibitemShut {NoStop}%
\bibitem [{\citenamefont {Solovov}\ \emph {et~al.}(2012)\citenamefont 
{Solovov}
  \emph {et~al.}}]{Solovov:2012}%
  \BibitemOpen
  \bibfield  {author} {\bibinfo {author} {\bibfnamefont {V.N.}\ 
\bibnamefont
  {Solovov}} \emph {et~al.} (\bibinfo {collaboration} {ZEPLIN-III
  Collaboration}),\ }\bibfield  {title} {\emph {\bibinfo {title} 
{Position
  reconstruction in a dual phase xenon scintillation detector}},\ 
}\href
  {\doibase 10.1109/TNS.2012.2221742} {\bibfield  {journal} {\bibinfo
  {journal} {IEEE Trans.\ Nucl.\ Sci.}\ }\textbf {\bibinfo {volume} 
{59}},\
  \bibinfo {pages} {3286--3293} (\bibinfo {year} {2012})}\BibitemShut 
{NoStop}%
\bibitem [{\citenamefont {Mei}(2011)}]{Mei:2011}%
  \BibitemOpen
  \bibfield  {author} {\bibinfo {author} {\bibfnamefont 
{Y}~\bibnamefont
  {Mei}},\ }\emph {\bibinfo {title} {Direct dark matter search with the
  {XENON100} experiment}},\ \href
  {https://scholarship.rice.edu/handle/1911/70350} {Ph.D. thesis},\ 
\bibinfo
  {school} {Rice University} (\bibinfo {year} {2011})\BibitemShut 
{NoStop}%
\bibitem [{\citenamefont {Akerib}\ \emph {et~al.}(in
  preparation{\natexlab{a}})\citenamefont {Akerib} \emph
  {et~al.}}]{Akerib:2015:comprehensive}%
  \BibitemOpen
  \bibfield  {author} {\bibinfo {author} {\bibfnamefont 
{D.}~\bibnamefont
  {Akerib}} \emph {et~al.} (\bibinfo {collaboration} {LUX 
Collaboration}),\
  }\href@noop {} {\emph {\bibinfo {title} {Calibration, event
  reconstruction, data analysis and limits calculation for the {LUX} 
dark
  matter experiment}},\ } (\bibinfo {year} {in
  preparation}{\natexlab{a}})\BibitemShut {NoStop}%
\bibitem [{\citenamefont {Kastens}\ \emph {et~al.}(2009)\citenamefont
  {Kastens}, \citenamefont {Cahn}, \citenamefont {Manzur},\ and\ 
\citenamefont
  {McKinsey}}]{Kastens:2009}%
  \BibitemOpen
  \bibfield  {author} {\bibinfo {author} {\bibfnamefont {L.W.}\ 
\bibnamefont
  {Kastens}}, \bibinfo {author} {\bibfnamefont {S.B.}\ \bibnamefont 
{Cahn}},
  \bibinfo {author} {\bibfnamefont {A.}~\bibnamefont {Manzur}}, \ and\ 
\bibinfo
  {author} {\bibfnamefont {D.N.}\ \bibnamefont {McKinsey}},\ }\bibfield
  {title} {\emph {\bibinfo {title} {Calibration of a liquid xenon 
detector
  with $^{83}\mathrm{Kr}{}^{m}$}},\ }\href {\doibase
  10.1103/PhysRevC.80.045809} {\bibfield  {journal} {\bibinfo  
{journal}
  {Phys.\ Rev.\ C}\ }\textbf {\bibinfo {volume} {80}},\ \bibinfo 
{pages}
  {045809} (\bibinfo {year} {2009})}\BibitemShut {NoStop}%
\bibitem [{\citenamefont {Manalaysay}\ \emph 
{et~al.}(2010)\citenamefont
  {Manalaysay}, \citenamefont {Marrod{\'a}n~Undagoitia}, \citenamefont 
{Askin},
  \citenamefont {Baudis}, \citenamefont {Behrens}, \citenamefont 
{Ferella},
  \citenamefont {Kish}, \citenamefont {Lebeda}, \citenamefont 
{Santorelli},
  \citenamefont {Vénos},\ and\ \citenamefont 
{Vollhardt}}]{Manalaysay:2010}%
  \BibitemOpen
  \bibfield  {author} {\bibinfo {author} {\bibfnamefont 
{A.}~\bibnamefont
  {Manalaysay}}, \bibinfo {author} {\bibfnamefont {T.}~\bibnamefont
  {Marrod{\'a}n~Undagoitia}}, \bibinfo {author} {\bibfnamefont
  {A.}~\bibnamefont {Askin}}, \bibinfo {author} {\bibfnamefont
  {L.}~\bibnamefont {Baudis}}, \bibinfo {author} {\bibfnamefont
  {A.}~\bibnamefont {Behrens}}, \bibinfo {author} {\bibfnamefont 
{A.~D.}\
  \bibnamefont {Ferella}}, \bibinfo {author} {\bibfnamefont 
{A.}~\bibnamefont
  {Kish}}, \bibinfo {author} {\bibfnamefont {O.}~\bibnamefont 
{Lebeda}},
  \bibinfo {author} {\bibfnamefont {R.}~\bibnamefont {Santorelli}}, 
\bibinfo
  {author} {\bibfnamefont {D.}~\bibnamefont {Vénos}}, \ and\ \bibinfo 
{author}
  {\bibfnamefont {A.}~\bibnamefont {Vollhardt}},\ }\bibfield  {title} 
{\emph
  {\bibinfo {title} {{Spatially uniform calibration of a liquid xenon 
detector
  at low energies using 83m-Kr}}},\ }\href {\doibase 10.1063/1.3436636}
  {\bibfield  {journal} {\bibinfo  {journal} {Rev.\ Sci.\ Instrum.}\ 
}\textbf
  {\bibinfo {volume} {81}},\ \bibinfo {pages} {073303} (\bibinfo {year}
  {2010})}\BibitemShut {NoStop}%
\bibitem [{\citenamefont {Akerib}\ \emph {et~al.}(in
  preparation{\natexlab{b}})\citenamefont {Akerib} \emph
  {et~al.}}]{Akerib:2016:dd}%
  \BibitemOpen
  \bibfield  {author} {\bibinfo {author} {\bibfnamefont {D.S.}\ 
\bibnamefont
  {Akerib}} \emph {et~al.} (\bibinfo {collaboration} {LUX 
Collaboration}),\
  }\href@noop {} {\emph {\bibinfo {title} {Low energy (0.7--74~{keV})
  nuclear recoil calibration of the {LUX} dark matter experiment using
  \mbox{{D}-{D}} neutron scattering kinematics}},\ } (\bibinfo {year} 
{in
  preparation}{\natexlab{b}})\BibitemShut {NoStop}%
\bibitem [{\citenamefont {Akerib}\ \emph
  {et~al.}(2016{\natexlab{a}})\citenamefont {Akerib} \emph
  {et~al.}}]{Akerib:2015:tritium}%
  \BibitemOpen
  \bibfield  {author} {\bibinfo {author} {\bibfnamefont {D.~S.}\ 
\bibnamefont
  {Akerib}} \emph {et~al.} (\bibinfo {collaboration} {LUX 
Collaboration}),\
  }\bibfield  {title} {\emph {\bibinfo {title} {Tritium calibration of 
the
  {LUX} dark matter experiment}},\ }\href {\doibase 
10.1103/PhysRevD.93.072009}
  {\bibfield  {journal} {\bibinfo  {journal} {Phys. Rev. D}\ }\textbf 
{\bibinfo
  {volume} {93}},\ \bibinfo {pages} {072009} (\bibinfo {year}
  {2016}{\natexlab{a}})}\BibitemShut {NoStop}%
\bibitem [{\citenamefont {Doke}\ \emph {et~al.}(2002)\citenamefont 
{Doke},
  \citenamefont {Hitachi}, \citenamefont {Kikuchi}, \citenamefont 
{Masuda},
  \citenamefont {Okada},\ and\ \citenamefont {Shibamura}}]{Doke:2002}%
  \BibitemOpen
  \bibfield  {author} {\bibinfo {author} {\bibfnamefont {Tadayoshi}\
  \bibnamefont {Doke}}, \bibinfo {author} {\bibfnamefont {Akira}\ 
\bibnamefont
  {Hitachi}}, \bibinfo {author} {\bibfnamefont {Jun}\ \bibnamefont 
{Kikuchi}},
  \bibinfo {author} {\bibfnamefont {Kimiaki}\ \bibnamefont {Masuda}}, 
\bibinfo
  {author} {\bibfnamefont {Hiroyuki}\ \bibnamefont {Okada}}, \ and\ 
\bibinfo
  {author} {\bibfnamefont {Eido}\ \bibnamefont {Shibamura}},\ 
}\bibfield
  {title} {\emph {\bibinfo {title} {Absolute scintillation yields in 
liquid
  argon and xenon for various particles}},\ }\href
  {http://stacks.iop.org/1347-4065/41/i=3R/a=1538} {\bibfield  
{journal}
  {\bibinfo  {journal} {Jpn.\ J.\ Appl.\ Phys.}\ }\textbf {\bibinfo 
{volume}
  {41}},\ \bibinfo {pages} {1538} (\bibinfo {year} {2002})}\BibitemShut
  {NoStop}%
\bibitem [{\citenamefont {Phelps}(2014)}]{Phelps:2014}%
  \BibitemOpen
  \bibfield  {author} {\bibinfo {author} {\bibfnamefont 
{P}~\bibnamefont
  {Phelps}},\ }\emph {\bibinfo {title} {The {LUX} dark matter 
experiment:
  detector performance and energy calibration}},\ \href
  {http://rave.ohiolink.edu/etdc/view?acc_num=case1404908222} {Ph.D. 
thesis},\
  \bibinfo  {school} {Case Western Reserve University} (\bibinfo {year}
  {2014})\BibitemShut {NoStop}%
\bibitem [{\citenamefont {Shutt}\ \emph {et~al.}(2007)\citenamefont 
{Shutt},
  \citenamefont {Dahl}, \citenamefont {Kwong}, \citenamefont 
{Bolozdynya},\
  and\ \citenamefont {Brusov}}]{Shutt:2007}%
  \BibitemOpen
  \bibfield  {author} {\bibinfo {author} {\bibfnamefont 
{T.}~\bibnamefont
  {Shutt}}, \bibinfo {author} {\bibfnamefont {C.E.}\ \bibnamefont 
{Dahl}},
  \bibinfo {author} {\bibfnamefont {J.}~\bibnamefont {Kwong}}, \bibinfo
  {author} {\bibfnamefont {A.}~\bibnamefont {Bolozdynya}}, \ and\ 
\bibinfo
  {author} {\bibfnamefont {P.}~\bibnamefont {Brusov}},\ }\bibfield  
{title}
  {\emph {\bibinfo {title} {{Performance and fundamental processes at 
low
  energy in a two-phase liquid xenon dark matter detector}}},\ }\href 
{\doibase
  10.1016/j.nuclphysbps.2007.08.140} {\bibfield  {journal} {\bibinfo  
{journal}
  {Nucl.\ Phys.\ Proc.\ Suppl.}\ }\textbf {\bibinfo {volume} {173}},\ 
\bibinfo
  {pages} {160--163} (\bibinfo {year} {2007})}\BibitemShut {NoStop}%
\bibitem [{\citenamefont {Malling}(2014)}]{Malling:2014}%
  \BibitemOpen
  \bibfield  {author} {\bibinfo {author} {\bibfnamefont {D.C.}\ 
\bibnamefont
  {Malling}},\ }\emph {\bibinfo {title} {Measurement and analysis of 
{WIMP}
  detection backgrounds, and characterization and performance of the 
large
  underground xenon dark matter search experiment}},\ \href
  {https://repository.library.brown.edu/studio/item/bdr:386168/} {Ph.D.
  thesis},\ \bibinfo  {school} {Brown University} (\bibinfo {year}
  {2014})\BibitemShut {NoStop}%
\bibitem [{\citenamefont {Verbus}\ \emph {et~al.}()\citenamefont 
{Verbus} \emph
  {et~al.}}]{Verbus:2015}%
  \BibitemOpen
  \bibfield  {author} {\bibinfo {author} {\bibfnamefont {J.R.}\ 
\bibnamefont
  {Verbus}} \emph {et~al.},\ }\href@noop {} {\emph {\bibinfo {title}
  {Proposed low-energy absolute calibration of nuclear recoils in a 
liquid
  xenon {TPC} using \mbox{{D}-{D}} neutron scattering kinematics}},\ 
}\bibinfo
  {note} {(in preparation)}\BibitemShut {NoStop}%
\bibitem [{\citenamefont {Lindhard}\ \emph {et~al.}(1963)\citenamefont
  {Lindhard}, \citenamefont {Nielsen},\ and\ \citenamefont
  {Scharff}}]{Lindhard:1963}%
  \BibitemOpen
  \bibfield  {author} {\bibinfo {author} {\bibfnamefont 
{J.}~\bibnamefont
  {Lindhard}}, \bibinfo {author} {\bibfnamefont {V.}~\bibnamefont 
{Nielsen}}, \
  and\ \bibinfo {author} {\bibfnamefont {M.}~\bibnamefont {Scharff}},\
  }\bibfield  {title} {\emph {\bibinfo {title} {{Integral equations
  governing radiation effects}}},\ }\href
  {http://www.sdu.dk/media/bibpdf/Bind\%2030-39/Bind/mfm-33-10.pdf} 
{\bibfield
  {journal} {\bibinfo  {journal} {Mat.\ Fys.\ Medd.\ K.\ Dan.\ Vidensk.
  Selsk.}\ }\textbf {\bibinfo {volume} {33}},\ \bibinfo {pages} {10} 
(\bibinfo
  {year} {1963})}\BibitemShut {NoStop}%
\bibitem [{\citenamefont {Thomas}\ and\ \citenamefont
  {Imel}(1987)}]{Thomas:1987}%
  \BibitemOpen
  \bibfield  {author} {\bibinfo {author} {\bibfnamefont 
{J.}~\bibnamefont
  {Thomas}}\ and\ \bibinfo {author} {\bibfnamefont {D.~A.}\ 
\bibnamefont
  {Imel}},\ }\bibfield  {title} {\emph {\bibinfo {title} 
{Recombination of
  electron-ion pairs in liquid argon and liquid xenon}},\ }\href 
{\doibase
  10.1103/PhysRevA.36.614} {\bibfield  {journal} {\bibinfo  {journal} 
{Phys.\
  Rev.\ A}\ }\textbf {\bibinfo {volume} {36}},\ \bibinfo {pages} 
{614--616}
  (\bibinfo {year} {1987})}\BibitemShut {NoStop}%
\bibitem [{\citenamefont {Hitachi}\ \emph {et~al.}(1992)\citenamefont
  {Hitachi}, \citenamefont {Doke},\ and\ \citenamefont
  {Mozumder}}]{Hitachi:1992}%
  \BibitemOpen
  \bibfield  {author} {\bibinfo {author} {\bibfnamefont 
{A.}~\bibnamefont
  {Hitachi}}, \bibinfo {author} {\bibfnamefont {T.}~\bibnamefont 
{Doke}}, \
  and\ \bibinfo {author} {\bibfnamefont {A.}~\bibnamefont {Mozumder}},\
  }\bibfield  {title} {\emph {\bibinfo {title} {Luminescence quenching 
in
  liquid argon under charged-particle impact: Relative scintillation 
yield at
  different linear energy transfers}},\ }\href {\doibase
  10.1103/PhysRevB.46.11463} {\bibfield  {journal} {\bibinfo  
{journal} {Phys.\
  Rev.\ B}\ }\textbf {\bibinfo {volume} {46}},\ \bibinfo {pages} 
{11463--11470}
  (\bibinfo {year} {1992})}\BibitemShut {NoStop}%
\bibitem [{\citenamefont {Mei}\ \emph {et~al.}(2008)\citenamefont 
{Mei},
  \citenamefont {Yin}, \citenamefont {Stonehill},\ and\ \citenamefont
  {Hime}}]{Mei:2008}%
  \BibitemOpen
  \bibfield  {author} {\bibinfo {author} {\bibfnamefont {D.-M.}\ 
\bibnamefont
  {Mei}}, \bibinfo {author} {\bibfnamefont {Z.-B.}\ \bibnamefont 
{Yin}},
  \bibinfo {author} {\bibfnamefont {L.C.}\ \bibnamefont {Stonehill}}, 
\ and\
  \bibinfo {author} {\bibfnamefont {A.}~\bibnamefont {Hime}},\ 
}\bibfield
  {title} {\emph {\bibinfo {title} {A model of nuclear recoil 
scintillation
  efficiency in noble liquids}},\ }\href {\doibase
  http://dx.doi.org/10.1016/j.astropartphys.2008.06.001} {\bibfield  
{journal}
  {\bibinfo  {journal} {Astropart.\ Phys.}\ }\textbf {\bibinfo 
{volume} {30}},\
  \bibinfo {pages} {12 -- 17} (\bibinfo {year} {2008})}\BibitemShut 
{NoStop}%
\bibitem [{\citenamefont {Lenardo}\ \emph {et~al.}(2015)\citenamefont
  {Lenardo}, \citenamefont {Kazkaz}, \citenamefont {Manalaysay}, 
\citenamefont
  {Mock}, \citenamefont {Szydagis},\ and\ \citenamefont
  {Tripathi}}]{Lenardo:2014}%
  \BibitemOpen
  \bibfield  {author} {\bibinfo {author} {\bibfnamefont 
{B.}~\bibnamefont
  {Lenardo}}, \bibinfo {author} {\bibfnamefont {K.}~\bibnamefont 
{Kazkaz}},
  \bibinfo {author} {\bibfnamefont {A.}~\bibnamefont {Manalaysay}}, 
\bibinfo
  {author} {\bibfnamefont {J.}~\bibnamefont {Mock}}, \bibinfo {author}
  {\bibfnamefont {M.}~\bibnamefont {Szydagis}}, \ and\ \bibinfo 
{author}
  {\bibfnamefont {M.}~\bibnamefont {Tripathi}},\ }\bibfield  {title} 
{\emph
  {\bibinfo {title} {A global analysis of light and charge yields in 
liquid
  xenon}},\ }\href {\doibase 10.1109/TNS.2015.2481322} {\bibfield  
{journal}
  {\bibinfo  {journal} {IEEE Trans.\ Nucl.\ Sci.}\ }\textbf {\bibinfo 
{volume}
  {62}},\ \bibinfo {pages} {3387} (\bibinfo {year} {2015})}\BibitemShut
  {NoStop}%
\bibitem [{\citenamefont {Bezrukov}\ \emph {et~al.}(2011)\citenamefont
  {Bezrukov}, \citenamefont {Kahlhoefer},\ and\ \citenamefont
  {Lindner}}]{Bezrukov:2011}%
  \BibitemOpen
  \bibfield  {author} {\bibinfo {author} {\bibfnamefont {Fedor}\ 
\bibnamefont
  {Bezrukov}}, \bibinfo {author} {\bibfnamefont {Felix}\ \bibnamefont
  {Kahlhoefer}}, \ and\ \bibinfo {author} {\bibfnamefont {Manfred}\
  \bibnamefont {Lindner}},\ }\bibfield  {title} {\emph {\bibinfo 
{title}
  {{Interplay between scintillation and ionization in liquid xenon 
Dark Matter
  searches}}},\ }\href {\doibase 10.1016/j.astropartphys.2011.06.008}
  {\bibfield  {journal} {\bibinfo  {journal} {Astropart.\ Phys.}\ 
}\textbf
  {\bibinfo {volume} {35}},\ \bibinfo {pages} {119--127} (\bibinfo 
{year}
  {2011})}\BibitemShut {NoStop}%
\bibitem [{\citenamefont {Akerib}\ \emph {et~al.}(2015)\citenamefont 
{Akerib}
  \emph {et~al.}}]{Akerib:2015:bg}%
  \BibitemOpen
  \bibfield  {author} {\bibinfo {author} {\bibfnamefont {D.S.}\ 
\bibnamefont
  {Akerib}} \emph {et~al.} (\bibinfo {collaboration} {LUX 
Collaboration}),\
  }\bibfield  {title} {\emph {\bibinfo {title} {Radiogenic and 
muon-induced
  backgrounds in the {LUX} dark matter detector}},\ }\href {\doibase
  http://dx.doi.org/10.1016/j.astropartphys.2014.07.009} {\bibfield  
{journal}
  {\bibinfo  {journal} {Astropart.\ Phys.}\ }\textbf {\bibinfo 
{volume} {62}},\
  \bibinfo {pages} {33 -- 46} (\bibinfo {year} {2015})}\BibitemShut 
{NoStop}%
\bibitem [{\citenamefont {Agostinelli}\ \emph 
{et~al.}(2003)\citenamefont
  {Agostinelli} \emph {et~al.}}]{Agostinelli:2003}%
  \BibitemOpen
  \bibfield  {author} {\bibinfo {author} {\bibfnamefont 
{S.}~\bibnamefont
  {Agostinelli}} \emph {et~al.} (\bibinfo {collaboration} {{GEANT4}
  Collaboration}),\ }\bibfield  {title} {\emph {\bibinfo {title} 
{Geant4---a
  simulation toolkit}},\ }\href {\doibase
  http://dx.doi.org/10.1016/S0168-9002(03)01368-8} {\bibfield  
{journal}
  {\bibinfo  {journal} {Nucl.\ Instrum.\ Methods Phys.\, Sect.\ A}\ 
}\textbf
  {\bibinfo {volume} {506}},\ \bibinfo {pages} {250--303} (\bibinfo 
{year}
  {2003})}\BibitemShut {NoStop}%
\bibitem [{\citenamefont {Szydagis}\ \emph {et~al.}(2011)\citenamefont
  {Szydagis}, \citenamefont {Barry}, \citenamefont {Kazkaz}, 
\citenamefont
  {Mock}, \citenamefont {Stolp}, \citenamefont {Sweany}, \citenamefont
  {Tripathi}, \citenamefont {Uvarov}, \citenamefont {Walsh},\ and\
  \citenamefont {Woods}}]{Szydagis:2011}%
  \BibitemOpen
  \bibfield  {author} {\bibinfo {author} {\bibfnamefont 
{M}~\bibnamefont
  {Szydagis}}, \bibinfo {author} {\bibfnamefont {N}~\bibnamefont 
{Barry}},
  \bibinfo {author} {\bibfnamefont {K}~\bibnamefont {Kazkaz}}, \bibinfo
  {author} {\bibfnamefont {J}~\bibnamefont {Mock}}, \bibinfo {author}
  {\bibfnamefont {D}~\bibnamefont {Stolp}}, \bibinfo {author} 
{\bibfnamefont
  {M}~\bibnamefont {Sweany}}, \bibinfo {author} {\bibfnamefont 
{M}~\bibnamefont
  {Tripathi}}, \bibinfo {author} {\bibfnamefont {S}~\bibnamefont 
{Uvarov}},
  \bibinfo {author} {\bibfnamefont {N}~\bibnamefont {Walsh}}, \ and\ 
\bibinfo
  {author} {\bibfnamefont {M}~\bibnamefont {Woods}},\ }\bibfield  
{title}
  {\emph {\bibinfo {title} {{NEST}: a comprehensive model for 
scintillation
  yield in liquid xenon}},\ }\href {\doibase 
10.1088/1748-0221/6/10/P10002}
  {\bibfield  {journal} {\bibinfo  {journal} {J.\ Inst.}\ }\textbf 
{\bibinfo
  {volume} {6}},\ \bibinfo {pages} {P10002} (\bibinfo {year}
  {2011})}\BibitemShut {NoStop}%
\bibitem [{\citenamefont {Szydagis}\ \emph {et~al.}(2013)\citenamefont
  {Szydagis}, \citenamefont {Fyhrie}, \citenamefont {Thorngren},\ and\
  \citenamefont {Tripathi}}]{Szydagis:2013}%
  \BibitemOpen
  \bibfield  {author} {\bibinfo {author} {\bibfnamefont 
{M}~\bibnamefont
  {Szydagis}}, \bibinfo {author} {\bibfnamefont {A}~\bibnamefont 
{Fyhrie}},
  \bibinfo {author} {\bibfnamefont {D}~\bibnamefont {Thorngren}}, \ 
and\
  \bibinfo {author} {\bibfnamefont {M}~\bibnamefont {Tripathi}},\ 
}\bibfield
  {title} {\emph {\bibinfo {title} {Enhancement of {NEST} capabilities 
for
  simulating low-energy recoils in liquid xenon}},\ }\href
  {http://stacks.iop.org/1748-0221/8/i=10/a=C10003} {\bibfield  
{journal}
  {\bibinfo  {journal} {J.\ Inst.}\ }\textbf {\bibinfo {volume} {8}},\ 
\bibinfo
  {pages} {C10003} (\bibinfo {year} {2013})}\BibitemShut {NoStop}%
\bibitem [{\citenamefont {Akerib}\ \emph {et~al.}(2012)\citenamefont 
{Akerib}
  \emph {et~al.}}]{Akerib:2012:luxsim}%
  \BibitemOpen
  \bibfield  {author} {\bibinfo {author} {\bibfnamefont 
{DS}~\bibnamefont
  {Akerib}} \emph {et~al.} (\bibinfo {collaboration} {LUX 
Collaboration}),\
  }\bibfield  {title} {\emph {\bibinfo {title} {{LUXSim}: A
  component-centric approach to low-background simulations}},\ }\href 
{\doibase
  10.1016/j.nima.2012.02.010"} {\bibfield  {journal} {\bibinfo  
{journal}
  {Nucl.\ Instrum.\ Methods Phys.\, Sect.\ A}\ }\textbf {\bibinfo 
{volume}
  {675}},\ \bibinfo {pages} {63--77} (\bibinfo {year} 
{2012})}\BibitemShut
  {NoStop}%
\bibitem [{\citenamefont {Lee}(2015)}]{Lee:2015}%
  \BibitemOpen
  \bibfield  {author} {\bibinfo {author} {\bibfnamefont 
{C}~\bibnamefont
  {Lee}},\ }\emph {\bibinfo {title} {Mitigation of backgrounds for the 
Large
  Underground Xenon dark matter experiment}},\ \href
  {http://rave.ohiolink.edu/etdc/view?acc_num=case1427482791} {Ph.D. 
thesis},\
  \bibinfo  {school} {Case Western Reserve University} (\bibinfo {year}
  {2015})\BibitemShut {NoStop}%
\bibitem [{\citenamefont {Chapman}(2014)}]{Chapman:2014}%
  \BibitemOpen
  \bibfield  {author} {\bibinfo {author} {\bibfnamefont {J.J.}\ 
\bibnamefont
  {Chapman}},\ }\emph {\bibinfo {title} {First {WIMP} search results 
from the
  {LUX} dark matter experiment}},\ \href
  {https://repository.library.brown.edu/studio/item/bdr:386289/} {Ph.D.
  thesis},\ \bibinfo  {school} {Brown University} (\bibinfo {year}
  {2014})\BibitemShut {NoStop}%
\bibitem [{\citenamefont {Cowan}\ \emph {et~al.}(2011)\citenamefont 
{Cowan},
  \citenamefont {Cranmer}, \citenamefont {Gross},\ and\ \citenamefont
  {Vitells}}]{Cowan:2011}%
  \BibitemOpen
  \bibfield  {author} {\bibinfo {author} {\bibfnamefont {Glen}\ 
\bibnamefont
  {Cowan}}, \bibinfo {author} {\bibfnamefont {Kyle}\ \bibnamefont 
{Cranmer}},
  \bibinfo {author} {\bibfnamefont {Eilam}\ \bibnamefont {Gross}}, \ 
and\
  \bibinfo {author} {\bibfnamefont {Ofer}\ \bibnamefont {Vitells}},\ 
}\bibfield
   {title} {\emph {\bibinfo {title} {{Asymptotic formulae for
  likelihood-based tests of new physics}}},\ }\href {\doibase
  10.1140/epjc/s10052-011-1554-0} {\bibfield  {journal} {\bibinfo  
{journal}
  {Eur.\ Phys.\ J. C}\ }\textbf {\bibinfo {volume} {71}},\ \bibinfo 
{pages}
  {1554} (\bibinfo {year} {2011})}\BibitemShut {NoStop}%
\bibitem [{\citenamefont {Moneta}\ \emph {et~al.}(2010)\citenamefont 
{Moneta},
  \citenamefont {Belasco}, \citenamefont {Cranmer}, \citenamefont 
{Kreiss},
  \citenamefont {Lazzaro}, \citenamefont {Piparo}, \citenamefont 
{Schott},
  \citenamefont {Verkerke},\ and\ \citenamefont {Wolf}}]{Moneta:2010}%
  \BibitemOpen
  \bibfield  {author} {\bibinfo {author} {\bibfnamefont 
{L.}~\bibnamefont
  {Moneta}}, \bibinfo {author} {\bibfnamefont {K.}~\bibnamefont 
{Belasco}},
  \bibinfo {author} {\bibfnamefont {K.~S.}\ \bibnamefont {Cranmer}}, 
\bibinfo
  {author} {\bibfnamefont {S.}~\bibnamefont {Kreiss}}, \bibinfo 
{author}
  {\bibfnamefont {A.}~\bibnamefont {Lazzaro}}, \bibinfo {author} 
{\bibfnamefont
  {D.}~\bibnamefont {Piparo}}, \bibinfo {author} {\bibfnamefont
  {G.}~\bibnamefont {Schott}}, \bibinfo {author} {\bibfnamefont
  {W.}~\bibnamefont {Verkerke}}, \ and\ \bibinfo {author} 
{\bibfnamefont
  {M.}~\bibnamefont {Wolf}},\ }\bibfield  {title} {\emph {\bibinfo 
{title}
  {{The RooStats project}}},\ }\href
  {http://pos.sissa.it/archive/conferences/093/057/ACAT2010_057.pdf} 
{\bibfield
   {journal} {\bibinfo  {journal} {Proc.\ Sci.}\ }\textbf {\bibinfo 
{volume}
  {ACAT2010}},\ \bibinfo {pages} {057} (\bibinfo {year} 
{2010})}\BibitemShut
  {NoStop}%
\bibitem [{\citenamefont {Cowan}\ \emph {et~al.}()\citenamefont 
{Cowan},
  \citenamefont {Cranmer}, \citenamefont {Gross},\ and\ \citenamefont
  {Vitells}}]{Cowan:2011:pcl}%
  \BibitemOpen
  \bibfield  {author} {\bibinfo {author} {\bibfnamefont {Glen}\ 
\bibnamefont
  {Cowan}}, \bibinfo {author} {\bibfnamefont {Kyle}\ \bibnamefont 
{Cranmer}},
  \bibinfo {author} {\bibfnamefont {Eilam}\ \bibnamefont {Gross}}, \ 
and\
  \bibinfo {author} {\bibfnamefont {Ofer}\ \bibnamefont {Vitells}},\ 
}\bibfield
   {title} {\emph {\bibinfo {title} {{Power-constrained limits}}},\
  }\href@noop {} {\ }\Eprint {http://arxiv.org/abs/1105.3166} 
{arXiv:1105.3166}
  \BibitemShut {NoStop}%
\bibitem [{\citenamefont {McCabe}(2010)}]{McCabe:2010}%
  \BibitemOpen
  \bibfield  {author} {\bibinfo {author} {\bibfnamefont {Christopher}\
  \bibnamefont {McCabe}},\ }\bibfield  {title} {\emph {\bibinfo {title}
  {Astrophysical uncertainties of dark matter direct detection 
experiments}},\
  }\href {\doibase 10.1103/PhysRevD.82.023530} {\bibfield  {journal} 
{\bibinfo
  {journal} {Phys. Rev. D}\ }\textbf {\bibinfo {volume} {82}},\ 
\bibinfo
  {pages} {023530} (\bibinfo {year} {2010})}\BibitemShut {NoStop}%
\bibitem [{\citenamefont {Akerib}\ \emph
  {et~al.}(2016{\natexlab{b}})\citenamefont {Akerib} \emph
  {et~al.}}]{Akerib:2016:sd}%
  \BibitemOpen
  \bibfield  {author} {\bibinfo {author} {\bibfnamefont {D.S.}\ 
\bibnamefont
  {Akerib}} \emph {et~al.} (\bibinfo {collaboration} {LUX 
Collaboration}),\
  }\bibfield  {title} {\emph {\bibinfo {title} {Results on the
  spin-dependent scattering of weakly interacting massive particles on 
nucleons
  from the run 3 data of the {LUX} experiment}},\ }\href {\doibase
  10.1103/PhysRevLett.116.161302} {\bibfield  {journal} {\bibinfo  
{journal}
  {Phys. Rev. Lett.}\ }\textbf {\bibinfo {volume} {116}},\ \bibinfo 
{pages}
  {161302} (\bibinfo {year} {2016}{\natexlab{b}})}\BibitemShut 
{NoStop}%
\bibitem [{\citenamefont {Agnese}\ \emph {et~al.}(2014)\citenamefont 
{Agnese}
  \emph {et~al.}}]{Scdms:2014}%
  \BibitemOpen
  \bibfield  {author} {\bibinfo {author} {\bibfnamefont 
{R.}~\bibnamefont
  {Agnese}} \emph {et~al.} (\bibinfo {collaboration} {SuperCDMS
  Collaboration}),\ }\bibfield  {title} {\emph {\bibinfo {title} 
{Search for
  low-mass weakly interacting massive particles with {SuperCDMS}}},\ 
}\href
  {\doibase 10.1103/PhysRevLett.112.241302} {\bibfield  {journal} 
{\bibinfo
  {journal} {Phys.\ Rev.\ Lett.}\ }\textbf {\bibinfo {volume} {112}},\ 
\bibinfo
  {eid} {241302} (\bibinfo {year} {2014})}\BibitemShut {NoStop}%
\bibitem [{\citenamefont {Agnese}\ \emph {et~al.}(2016)\citenamefont 
{Agnese}
  \emph {et~al.}}]{Cdmslite:2015}%
  \BibitemOpen
  \bibfield  {author} {\bibinfo {author} {\bibfnamefont 
{R.}~\bibnamefont
  {Agnese}} \emph {et~al.} (\bibinfo {collaboration} {SuperCDMS
  Collaboration}),\ }\bibfield  {title} {\emph {\bibinfo {title} {New
  results from the search for low-mass weakly interacting massive 
particles
  with the {CDMS} low ionization threshold experiment}},\ }\href 
{\doibase
  10.1103/PhysRevLett.116.071301} {\bibfield  {journal} {\bibinfo  
{journal}
  {Phys. Rev. Lett.}\ }\textbf {\bibinfo {volume} {116}},\ \bibinfo 
{pages}
  {071301} (\bibinfo {year} {2016})}\BibitemShut {NoStop}%
\bibitem [{\citenamefont {Aprile}\ \emph {et~al.}(2012)\citenamefont 
{Aprile}
  \emph {et~al.}}]{Xenon100:2012}%
  \BibitemOpen
  \bibfield  {author} {\bibinfo {author} {\bibfnamefont 
{E.}~\bibnamefont
  {Aprile}} \emph {et~al.} (\bibinfo {collaboration} {XENON100
  Collaboration}),\ }\bibfield  {title} {\emph {\bibinfo {title} {Dark
  matter results from 225 live days of {XENON100} data}},\ }\href 
{\doibase
  10.1103/PhysRevLett.109.181301} {\bibfield  {journal} {\bibinfo  
{journal}
  {Phys.\ Rev.\ Lett.}\ }\textbf {\bibinfo {volume} {109}},\ \bibinfo 
{eid}
  {181301} (\bibinfo {year} {2012})}\BibitemShut {NoStop}%
\bibitem [{\citenamefont {Agnes}\ \emph {et~al.}(2016)\citenamefont 
{Agnes}
  \emph {et~al.}}]{Darkside:2015}%
  \BibitemOpen
  \bibfield  {author} {\bibinfo {author} {\bibfnamefont 
{P.}~\bibnamefont
  {Agnes}} \emph {et~al.} (\bibinfo {collaboration} {DarkSide 
Collaboration}),\
  }\bibfield  {title} {\emph {\bibinfo {title} {Results from the first 
use
  of low radioactivity argon in a dark matter search}},\ }\href 
{\doibase
  10.1103/PhysRevD.93.081101} {\bibfield  {journal} {\bibinfo  
{journal} {Phys.
  Rev. D}\ }\textbf {\bibinfo {volume} {93}},\ \bibinfo {pages} 
{081101}
  (\bibinfo {year} {2016})}\BibitemShut {NoStop}%
\bibitem [{\citenamefont {Xiao}\ \emph {et~al.}(2015)\citenamefont 
{Xiao} \emph
  {et~al.}}]{Pandax:2015}%
  \BibitemOpen
  \bibfield  {author} {\bibinfo {author} {\bibfnamefont {Xiang}\ 
\bibnamefont
  {Xiao}} \emph {et~al.} (\bibinfo {collaboration} {PandaX 
Collaboration}),\
  }\bibfield  {title} {\emph {\bibinfo {title} {Low-mass dark matter 
search
  results from full exposure of the {PandaX-I} experiment}},\ }\href 
{\doibase
  10.1103/PhysRevD.92.052004} {\bibfield  {journal} {\bibinfo  
{journal} {Phys.
  Rev. D}\ }\textbf {\bibinfo {volume} {92}},\ \bibinfo {pages} 
{052004}
  (\bibinfo {year} {2015})}\BibitemShut {NoStop}%
\bibitem [{\citenamefont {Billard}\ \emph {et~al.}(2014)\citenamefont
  {Billard}, \citenamefont {Figueroa-Feliciano},\ and\ \citenamefont
  {Strigari}}]{Billard:2013}%
  \BibitemOpen
  \bibfield  {author} {\bibinfo {author} {\bibfnamefont 
{J.}~\bibnamefont
  {Billard}}, \bibinfo {author} {\bibfnamefont {E.}~\bibnamefont
  {Figueroa-Feliciano}}, \ and\ \bibinfo {author} {\bibfnamefont
  {L.}~\bibnamefont {Strigari}},\ }\bibfield  {title} {\emph {\bibinfo
  {title} {Implication of neutrino backgrounds on the reach of next 
generation
  dark matter direct detection experiments}},\ }\href {\doibase
  10.1103/PhysRevD.89.023524} {\bibfield  {journal} {\bibinfo  
{journal}
  {Phys.\ Rev.\ D}\ }\textbf {\bibinfo {volume} {89}},\ \bibinfo 
{pages}
  {023524} (\bibinfo {year} {2014})}\BibitemShut {NoStop}%
\bibitem [{\citenamefont {Akerib}\ \emph {et~al.}()\citenamefont 
{Akerib} \emph
  {et~al.}}]{Akerib:2015:cdr}%
  \BibitemOpen
  \bibfield  {author} {\bibinfo {author} {\bibfnamefont {D.S.}\ 
\bibnamefont
  {Akerib}} \emph {et~al.} (\bibinfo {collaboration} {LZ 
Collaboration}),\
  }\bibfield  {title} {\emph {\bibinfo {title} {{LUX-ZEPLIN conceptual
  design report}}},\ }\href@noop {} {\ }\Eprint
  {http://arxiv.org/abs/1509.02910v2} {arXiv:1509.02910v2} \BibitemShut
  {NoStop}%
\end{thebibliography}%

\end{document}